\documentclass[%
aip,
amsmath,amssymb,
preprint,
]{revtex4-1}

\usepackage{graphicx}
\usepackage{bm}
\usepackage[normalem]{ulem}
\usepackage{hyperref}
\hypersetup{colorlinks=true,urlcolor=blue,citecolor=blue,linkcolor=blue}
\usepackage[utf8]{inputenc}
\usepackage[T1]{fontenc}
\usepackage{mathptmx}
\usepackage{etoolbox}
\usepackage{subcaption}
\usepackage{xcolor}
\def\bx{{\bf x}}

\makeatletter
\def\@email#1#2{%
 \endgroup
 \patchcmd{\titleblock@produce}
  {\frontmatter@RRAPformat}
  {\frontmatter@RRAPformat{\produce@RRAP{*#1\href{mailto:#2}{#2}}}\frontmatter@RRAPformat}
  {}{}
}%

\makeatother
\usepackage{soul,xcolor}

\begin{document}

\setstcolor{red}

\title{Estimating solvation free energies with Boltzmann generators} 
\author{Maximilian Schebek}
\address{Department of Physics, Freie Universit\"at Berlin,
14195 Berlin, Germany}
\author{Nikolas M. Frob{\"o}se}
\author{Bettina G.~Keller}
\address{Department of Biology, Chemistry and Pharmacy, Freie Universit\"at Berlin,
14195 Berlin, Germany}
\author{Jutta Rogal}
\address{Initiative for Computational Catalysis, Flatiron Institute, New York, New York 10010, USA}

\date{\today}

\begin{abstract}
Accurate calculations of solvation free energies remain a central challenge in molecular simulations, often requiring extensive sampling and numerous alchemical intermediates to ensure sufficient overlap between phase-space distributions of a solute in the gas phase and in solution. Here, we introduce a computational framework based on normalizing flows that directly maps solvent configurations between solutes of different sizes,
and compare the accuracy and efficiency to conventional free energy estimates.
For a Lennard-Jones solvent, we demonstrate that this approach yields acceptable accuracy in estimating free energy differences for challenging transformations, such as solute growth or increased solute–solute separation, which typically demand multiple intermediate simulation steps along the transformation. Analysis of radial distribution functions indicates that the flow generates physically meaningful solvent rearrangements, substantially enhancing configurational overlap between states in configuration space. These results suggest flow-based models as a promising alternative to traditional free energy estimation methods.
\end{abstract}

\maketitle

\section{Introduction}
Transfering a molecule from the gas phase into a solvent is a complex molecular process.~\cite{kronberg2016hydrophobic, grdadolnik2017origin, pezzotti2022spectroscopic} 
Solvent interactions have to break and reorganize to form a cavity large enough to accommodate the molecule. 
Furthermore, the conformation of the molecule rearranges to maximize its free energy in the solvent, while new solute-solvent interactions form. 
In addition to enthalpic changes, solvation processes usually involve a considerable entropy change, both for the solvent and the solute.

The solvation free energy $\Delta F_{\mathrm{solv}}(A) = F_{\mathrm{sol}}(A)-F_{\mathrm{gas}}(A)$ is the reversible work required to transfer a molecule $A$ from the gas phase into a solvent, where $F_{\mathrm{gas}}(A)$ is the Helmholtz free energy of $A$ in the gas phase and $F_{\mathrm{sol}}(A)$  in the solvent, respectively.  
It measures how favourable ($\Delta F_{\mathrm{solv}}(A)<0$) or unfavourable ($\Delta F_{\mathrm{solv}}(A)>0$) this transfer is.
Solvation free energies are essential for understanding and predicting hydrophobicity, that is, the tendency of a solute to accumulate in non-polar surroundings rather than in water.
This tendency governs several  molecular phenomena, such as aggregation in aqueous solution, membrane permeability, protein folding, protein-ligand binding or, more general, host-guest-binding. 
Moreover solvation free-energies connect molecular simulations to experimentally accessible observables, including solubilities, partition coefficients and hydration enthalpies,~\cite{duarte2017approaches, bannan2016calculating} which makes them a critical benchmark for validating the underlying potential energy function.~\cite{mobley2009small, mobley2012alchemical, kelly2020alchemical}

Absolute solvation free energies can, in principle, be estimated from molecular simulations via alchemical free energy perturbation (FEP).~\cite{zwanzig_fep}
However, because solvation involves significant and concerted rearrangements of many solvent molecules, the initial (gas-phase) and final (fully solvated) states share almost no phase-space overlap, making direct alchemical FEP extremely difficult to converge.~\cite{shirts2007alchemical, mey2020best}
Relative solvation free energies, by contrast, are a practical and efficient way to obtain quantitative rankings of solvation free energies for different solutes $A$, $B$, $C$... .
This approach relies on a thermodynamic cycle, in which the calculation of the absolute solvation free energies $\Delta F_{\mathrm{solv}}(A)$ and $\Delta F_{\mathrm{solv}}(B)$ can be circumvented, and the relative solvation free energy of molecules $A$ and $B$
\begin{align}
    \Delta \Delta F_{\mathrm{solv}} = \Delta F_{\mathrm{solv}}(A)- \Delta F_{\mathrm{solv}}(B) = \Delta F_{\rm sol}^{A\rightarrow B} -\Delta F_{\rm gas}^{A\rightarrow B}
\label{eq:relativeSolvationFreeEnergy}    
\end{align}
is instead obtained from two alchemical transformations: transforming $A$ into $B$ in gas  phase ($\Delta F_{\rm gas}^{A\rightarrow B}$) and in solution ($\Delta F_{\rm sol}^{A\rightarrow B}$), respectively.
The transformation in the gas phase, $\Delta F_{\rm gas}^{A\rightarrow B}$, only involves the degrees of freedom of the solute and is relatively easy to converge.
In contrast, the transformation in solution, $\Delta F_{\rm sol}^{A\rightarrow B}$, requires large solvent rearrangements, which reduce phase-space overlap and make the corresponding free-energy calculations considerably harder to converge.
However, if $A$ and $B$ are sufficiently similar, $\Delta F_{\rm sol}^{A\rightarrow B}$ still converges significantly faster than the absolute solvation free energies $\Delta F_{\mathrm{solv}}(A)$ and $\Delta F_{\mathrm{solv}}(B)$, because in the alchemical transformation the solvent cavity is already formed and the orientations of solvent molecules in the first solvation shell remain similar.

Nonetheless, the numerical challenges involved in estimating alchemical free-energy changes in solution have motivated the development of a wide range of computational strategies,~\cite{chipot2007free, stoltz2010free, mey2020best, chipot2023free} which can broadly be classified into three categories. 
First, in one-step-perturbation approaches,~\cite{christ2007enveloping} an artificial reference Hamiltonian is constructed that has substantial phase-space overlap with both the initial state and the final state. 
By sampling the reference Hamiltonian, the free energy difference to these two states can directly be calculated,
and using the fact that the free energy is a state function, it directly yields the free-energy difference between initial and final state. 
Second, in multistaged approaches, a sequence of intermediate Hamiltonians is introduced that smoothly interpolate between the two endstates. 
These ensure sufficient phase-space overlap between neighboring states, such that their free-energy differences can be estimated reliably via direct FEP. 
The multistate Bennett acceptance ratio~\cite{Bennett1976-ic, Shirts2008-mbar} (MBAR) further improves statistical efficiency by optimally combining information from all simulated states, including non-neighboring ones.
Finally, targeted free energy perturbation (TFEP)~\cite{tfep} provides an alternative to mitigate the overlap problem by defining a suitable map on configuration space  
that transforms the configurations of the initial state to increase overlap with the final state.
However, constructing a suitable, invertible map by hand is non-trivial for any but the simplest systems which has hampered the widespread application of TFEP.

As an alternative, recent years have seen an increasing interest in leveraging the expressive power of neural networks to model high-dimensional probability distributions for enhancing free energy calculations. A promising class of neural free energy estimators is based on Boltzmann Generators~\cite{Noe2019} and learned free energy perturbation (LFEP),~\cite{wirnsberger_lfep} which leverage invertible transformations to enable exact likelihood calculations. These methods remove the need to handcraft a suitable mapping for TFEP-based approaches by parametrizing it as a learnable transformation,  making TFEP a  practical and powerful tool. This idea has been applied using both discrete and continuous normalizing flows,~\cite{papamakarios_flow} and has been shown to accelerate free energy estimations across the natural sciences, with applications ranging from proteins and small molecules~\cite{Rizzi2021, Rizzi2023,Olehnovics2024} to condensed phase systems such as liquids,~\cite{coretti_learning, wirnsberger_lfep} atomic solids,~\cite{Wirnsberger_2022, ahmad_free_2022, Wirnsberger_2023,schebek2024efficient, schebek2025scalable} and molecular crystals.~\cite{Olehnovics2025} Neural free energy estimators based on diffusion models~\cite{ho2020ddpm} have also been developed,~\cite{mate2024neural, mate2025solvation} although they often require extensive problem-specific preconditioning. Most recently, also non-equilibrium neural free energy estimators have been proposed.~\cite{he2025feat}

In this work, we focus on the Boltzmann Generator (BG) approach and show how it can be applied to the calculation of solvation free energies. Using a system of two solute particles in a solvent bath modeled with Lennard-Jones interactions as an example, we investigate chemically relevant processes that are difficult to capture with traditional free-energy estimators, including increasing the size of one of the solutes and evaluating the free-energy change associated with varying the distance between the two solute particles.

\section{Free energy estimators}
In the canonical ensemble ($NVT$), where the number of particles $N$, the temperature $T$, and the volume $V$ of the simulation box are kept constant, the equilibrium distribution is given by the Boltzmann distribution,
\begin{equation}
    \mu(\bx) = \frac{e^{-u(\bx)}}{Z}\quad,
\end{equation}
where \(u(\bx) = \beta U(\bx)\) is the reduced potential energy of the configuration $\bx\in\mathbb{R}^{3N}$, \(\beta = 1/(k_B T)\), and
\begin{equation}\label{eq:partition}
Z = \int d\bx\, e^{-u(\bx)}
\end{equation}
is the configurational partition function. The configurational part of the Helmholtz free energy $F$, commonly reported as dimensionless quantity $f = \beta F$, can be obtained from the partition function according to
\begin{equation}
f = - \ln Z\quad.
\end{equation}
Although, in principle, the partition function in Eq.~\eqref{eq:partition} can be expressed as a canonical ensemble average, the resulting estimator is dominated by high-energy configurations and is therefore impractical for direct computation. In practice, methods typically focus on computing free-energy differences between states  
for which reliable estimators are available.

\subsection{FEP and Multistate Bennett Acceptance Ratio}
FEP~\cite{zwanzig_fep} is an importance sampling-based method for computing the free energy difference between two states with  equilibrium distributions $q$ and $p$. Using configurations sampled from  \(q\), typically obtained through molecular dynamics (MD) or Monte Carlo (MC) simulations, FEP allows to compute the free energy difference $\Delta f_{qp} = \beta_p F_{p} - \beta_q F_q $  from the Zwanzig identity  as
\begin{equation}
\Delta f_{qp} = - \ln \mathbb{E}_{\bx \sim q}\Big[e^{-(u_p(\bx) - u_q(\bx))}\Big]\quad,
\end{equation}
where \(\mathbb{E}_{\bx \sim q}[\cdot]\) denotes the expectation over the equilibrium ensemble sampled from \(q\). The reduced potentials of $q$ and $p$ are defined by $u_q$ and $u_p$, respectively.   While formally exact, FEP is extremely sensitive to poor phase-space overlap between the configurations of $q$ and $p$, which can lead to large variance and slow convergence, or even make a free energy calculation impossible to perform numerically.

A straightforward solution to the overlap problem is the insertion of $K$ intermediate states along an appropriate order parameter and the stepwise calculation of free energy differences along this path. This procedure is adopted in MBAR,~\cite{Shirts2008-mbar} which provides a statistically optimal framework for combining data from all intermediate equilibrium ensembles to obtain the relative free energy of state $i$ via
\begin{equation} \label{eq:mbar}
    f_i= - \ln \sum_{j=1}^{K} \sum_{n=1}^{N_j} \frac{\exp[-u_i(\bx_{jn})]}{\sum_{k=1}^{K} N_k \exp[f_k - u_k(\bx_{jn})]} \quad ,
\end{equation}
where $N_j$ is the number of samples collected for state $j$. While intermediate states improve convergence, they come at the expense of additional MD or MC simulations, which may not always yield physically meaningful information. Furthermore, defining a suitable order parameter is not always trivial, and finding an effective balance between convergence and computational cost often requires tedious trial-and-error procedures.

TFEP~\cite{tfep} offers an alternative to introducing intermediate states by defining an invertible mapping $\mathcal{M} :\mathbb{R}^{3N} \rightarrow \mathbb{R}^{3N}$ on configuration space between the two states. By transforming configurations sampled from $q$, the mapping generates a new distribution $q'$ that shares greater overlap with $p$, which allows to compute the free energy difference between $q$ and $p$.
However, while TFEP has the potential to entirely remove the need for intermediate states, or at least greatly reduce their number, identifying a suitable transformation has long been infeasible for all but the simplest systems.

\subsection{Boltzmann Generators and Learned Free Energy Perturbation}
Boltzmann Generators (BG)~\cite{Noe2019} and learned free energy perturbation (LFEP)~\cite{wirnsberger_lfep} address the challenge of defining a suitable mapping in TFEP by learning the corresponding transformation. In particular, they leverage normalizing flows~\cite{papamakarios_flow} to learn an invertible mapping \(f_\theta\) that generates  a new distribution \( q_\theta\) with improved overlap with the target distribution \(p\) by transforming configurations \(\bx \sim q\), sampled from a base distribution $q$, via  $\bx' = f_\theta(\bx)$.  

The invertibility of \(f_\theta\) allows the density of the generated distribution $p_\theta$ to be computed analytically using the change-of-variables formula:
\begin{equation}
q_\theta(\bx') = q
(\bx) \, \Big|\det \frac{\partial f_\theta(\bx)}{\partial x}\Big|^{-1}\quad .
\end{equation}
In principle, there are two different ways of implementing the mapping using normalizing flows. In this work, we focus on coupling flows,~\cite{dinh_nice_2014, dinh2017density} a specific architecture with discrete transformations and easily computable Jacobians. This is achieved by partitioning the input into two channels and updating one conditioned on the other, producing a triangular Jacobian that can be evaluated analytically. Using coupling flows, the trainable parameters $\theta$ can be optimized by minimizing the reverse Kullback-Leibler (KL) divergence between 
$p$ and 
$q_\theta$, $D_\text{KL}(q_\theta \| p)$, an approach often referred to as \emph{energy-based} training. In contrast, continuous flow architectures entail alternative training paradigms and typically require samples from the target distribution for training.~\cite{lipman2022flow}

The exact likelihood property of flows allows to reweight the generated distribution to the target distribution using the importance weights $w$, which can be defined as
\begin{equation}
w(\bx) = e^{u_q(\bx) - u_p(f_\theta(\bx)) +\log|\det J_{f_\theta}(\bx)| } \quad,
\end{equation}
such that $w(\bx) \propto  p(f_\theta(\bx)) / q_\theta(f_\theta(\bx))$. Within TFEP, the free energy difference between base and target is obtained from the trained flow via 
\begin{equation}\label{eq:tfep}
\Delta f_{qp} = -\log \mathbb{E}_{\bx \sim q}\bigl[w(\bx)\bigr] \quad.
\end{equation}
The weights can be further utilized to define the Kish effective sample size~\cite{Kish1965-es} via
\begin{equation}\label{eq:ess_kish}
    \text{ESS} = \frac{\bigl[\sum_i w(\bx_i)\bigr]^2}{\sum_i [w(\bx_i)]^2} \quad,
\end{equation}
which is a useful diagnostic of flow performance as it provides an approximate measure of how many uncorrelated samples would yield a Monte Carlo estimator of comparable statistical quality.

\section{Experiments}
\begin{figure}[t]
\includegraphics[width=1\textwidth]{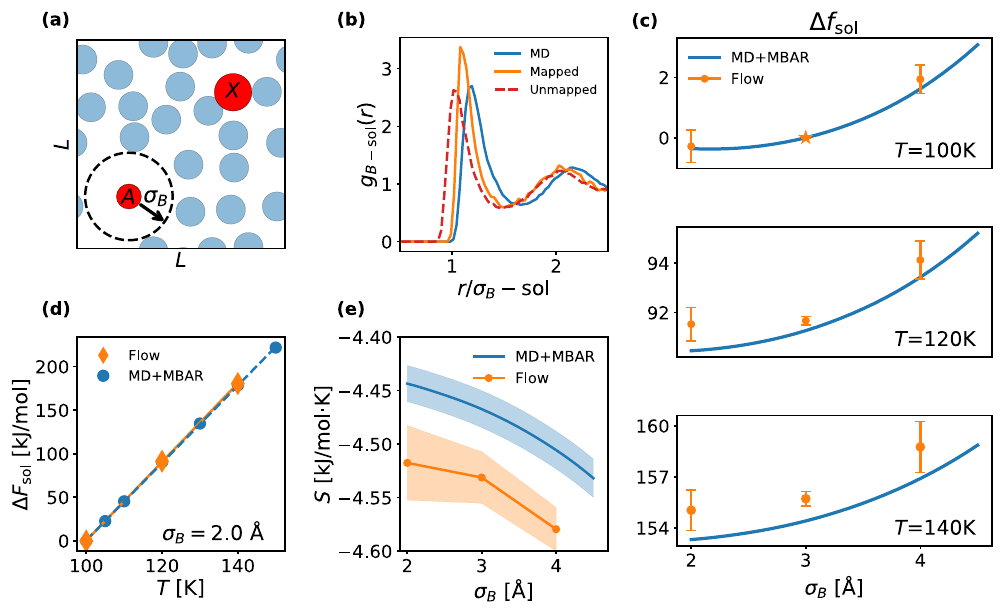}
\caption{\label{fig:exp_grow} (a) Illustration of the solvation experiment showing solvents (blue) and solutes $A$ and $X$ (red), with particle $A$ changing to $B$ by varying the radius across the simulations. (b) Radial distribution function  $g_{B-\mathrm{sol}}(r)$ between solute $B$ and the solvent particles at 100 K. The orange line (mapped) is obtained by applying the flow to solvent configurations from the $\sigma_B = 2.0~\text{\AA}$ simulation before inserting a $\sigma_B = 4.0~\text{\AA}$ solute, while the red dashed line (unmapped) uses the same solvent configurations without transformation prior to solute insertion. (c) Reduced free energy differences relative to the base state with $\sigma_B=3.0~\text{\AA}$ and $T=100~\text{K}$ (marked with a star symbol). (d) Change in $\Delta F$ for $\sigma_B=2.0~$\AA~as a function of temperature together with a linear regression of the data points. (e) Entropy obtained  as the slope of the linear fits in (d) for various $\sigma_B$-values.
}
\end{figure}

We consider two distinct solvation scenarios involving two solutes, $A$ and $X$, to demonstrate the applicability of the BG approach and compare its performance to traditional MD+MBAR simulations. 
In the first scenario, solute $A$ is transformed into a solute $B$ by systematically increasing its radius, while in addition also the temperature is varied (see Figure~\ref{fig:exp_grow} (a)). Thus, we calculate the free-energy difference in solution $\Delta F_{\rm sol}^{A\rightarrow B}$ from Eq.~\eqref{eq:relativeSolvationFreeEnergy}, while the free-energy difference in the gas phase, $\Delta F_{\rm gas}^{A\rightarrow B}$, is zero. 
The second scenario investigates the free energy difference upon varying the solute-solute distance $d_{AX}$ (see Figure \ref{fig:distance} (a)).
All simulations are performed for a cubic system at fixed volume with \(N = 100\) particles, including the two solute particles with fixed positions. While the solvent-solvent and solvent-solute interactions are modeled by a Lennard-Jones (LJ) potential, the solutes do not interact with each other.
Details regarding the LJ potential and simulation setup can be found in the supplementary information (SI).

For the MD+MBAR approach, intermediate states need to be sampled along the order parameter, i.e.,  the solute radius as well as the temperature in the first experiment, and the distance between the solutes in the second experiment. For each intermediate state, $S_{\rm MBAR }$ samples are collected using MD, which serve as an input to the MBAR estimator in Eq.~\eqref{eq:mbar}. For the flow approach, $S^{\rm train}_{\rm BG}$ samples are collected at a single base state, and mappings are learned to transform the solvent configurations to better match the target solute configurations. Free energy differences to the base state are obtained by generating an ensemble of  $S^{\rm eval}_{\rm BG}$ configurations in the target state and evaluating Eq.~\eqref{eq:tfep}. To allow for a fair comparison between the two approaches, we set  $S^{\rm eval}_{\rm BG}=S_{\rm MBAR }=1000$ for all  calculations. 

The flow model is implemented using a coupling architecture~\cite{dinh_nice_2014, dinh2017density} in which a subset of coordinates of a given solvent particles is updated conditioned on the remaining coordinates of the same atom. Since we did not observe any improvement in performance when including information from the surrounding solvent particles in the conditioning network, 
we employ this simpler, non-interacting formulation throughout. Computational details regarding the MD simulations and training of the flow model are given in the SI.

\subsection{Solute size dependent free energy}
In the first experiment, solute $A$, defined by a radius $\sigma_A$,  is transformed into solute $B$ with radius $\sigma_B$. The size of $B$ is varied by adjusting the Lennard-Jones interaction radius over the range $\sigma_B = 2.0$--$4.5$~\AA, and the corresponding free energy is also evaluated as a function of temperature over the range $T = 100$--$140$~K. The radius of the second solute is set to $\sigma_X =3.0$~\AA~.

The base distribution for the flow model is sampled at  $\sigma_A = 3.0$~\AA, $T^0 = 100~\mathrm{K}$ and all free energy differences are reported relative to this base state.
The reduced free energy differences as a function of $\sigma_B$ as obtained from MD+MBAR and the flow model are shown in Figure~\ref{fig:exp_grow}(c) for three different temperatures. 
At the reference temperature (top graph, $T=100$~K), the flow data shows excellent agreement with MBAR results. Most notably, even small free energy differences of less than 1~kJ/mol are accurately resolved. For higher temperatures, the flow needs to learn both a mapping of the solvent configuration with changing solute size $\sigma_B$ as well as temperature $T$. Over the investigated range, the free energy increases significantly with temperature while the change with solute size is much smaller and comparable at different $T$, as shown in the middle and bottom graph of Figure~\ref{fig:exp_grow}(c).
At $T=120$~K and 140~K, the overall agreement remains very good. The flow accurately reproduces the qualitative trends and remains consistent with MBAR results across all $T$ and $\sigma_B$, with the largest deviations being less than 2~kJ/mol.  Nevertheless, minor deviations are observed for these temperatures, consistent with very low observed sampling efficiencies (see Eq.~\eqref{eq:ess_kish}) below 1\%. Therefore, these results highlight the inherent difficulty of mapping equilibrium distributions of liquids onto each other, even when the mapping is only between different temperatures. This finding aligns with previous studies, which also report very low efficiencies when mapping distributions of different potential energy functions, even in cases where the potentials are very similar~\cite{coretti_learning} (see also Sec.~\ref{sec:flow_vs_mbar}). Importantly, these observations do not indicate that the estimator itself is biased, but rather that very small effective sample sizes can lead to biases in practice.

To illustrate the ability of the flow model to capture structural changes in the solvation shell around the solute, we compare radial distribution functions (RDFs) between solute particle $B$ with $\sigma_B = 4.0$~\AA~ and the solvent particles at $T=100$~K. The MD results (blue curve in Fig.~\ref{fig:exp_grow}(b)) show the clear signature of the first and second solvation shell surrounding particle $B$.
In addition, RDFs with and without applying the flow mapping are shown.
In the unmapped case (red dashed curve in Fig.~\ref{fig:exp_grow}(b)), the RDF is computed by taking the unaltered solvent configurations from the base state ($\sigma_B = \sigma_A = 3.0$~\AA) and increasing the solute radius to the target value. In the mapped case (orange curve in Fig.~\ref{fig:exp_grow}(b)), we start from the same solvent configurations but apply the flow mapping to propose updated solvent  configurations consistent with the larger solute. While the RDF generated by the flow does not perfectly match the target MD data, it  significantly increases the overlap as compared to the unmapped RDF  and highlights the flow’s ability to propose physically meaningful configurational updates. Although the RDF itself is purely structural, the deviation in the unmapped case indicates that a direct free energy estimate between the two states using FEP would be challenging, while the flow can improve the convergence of such estimates (see Sec.~\ref{sec:flow_vs_mbar} for a more detailed analysis).

A change in the size of the solute also affects the entropy $S$ of the system. With $S=-(\frac{\partial F}{\partial T})_V$, the 
entropy can be determined as the slope of $\Delta F$ as a function of temperature, as shown Figure~\ref{fig:exp_grow}(d) for $\sigma_B = 2.0$~\AA, by performing linear regression on the computed free energy values. 
The corresponding change in entropy as a function of $\sigma_B$ is illustrated in Figure~\ref{fig:exp_grow}(e). Both the MBAR and flow results exhibit the same overall trend, that is a decrease in the entropy for larger solute sizes, which is expected as a larger solute effectively reduces the accessible volume for the the solvent particles.
The slight offset between the entropy values obtained with the two approaches is less than $1.5\%$ and 
within their uncertainties, MBAR and flow estimates are consistent in their prediction of the entropy.

\subsection{Solute distance dependent free energy}
\begin{figure}
\includegraphics[width=1\textwidth]{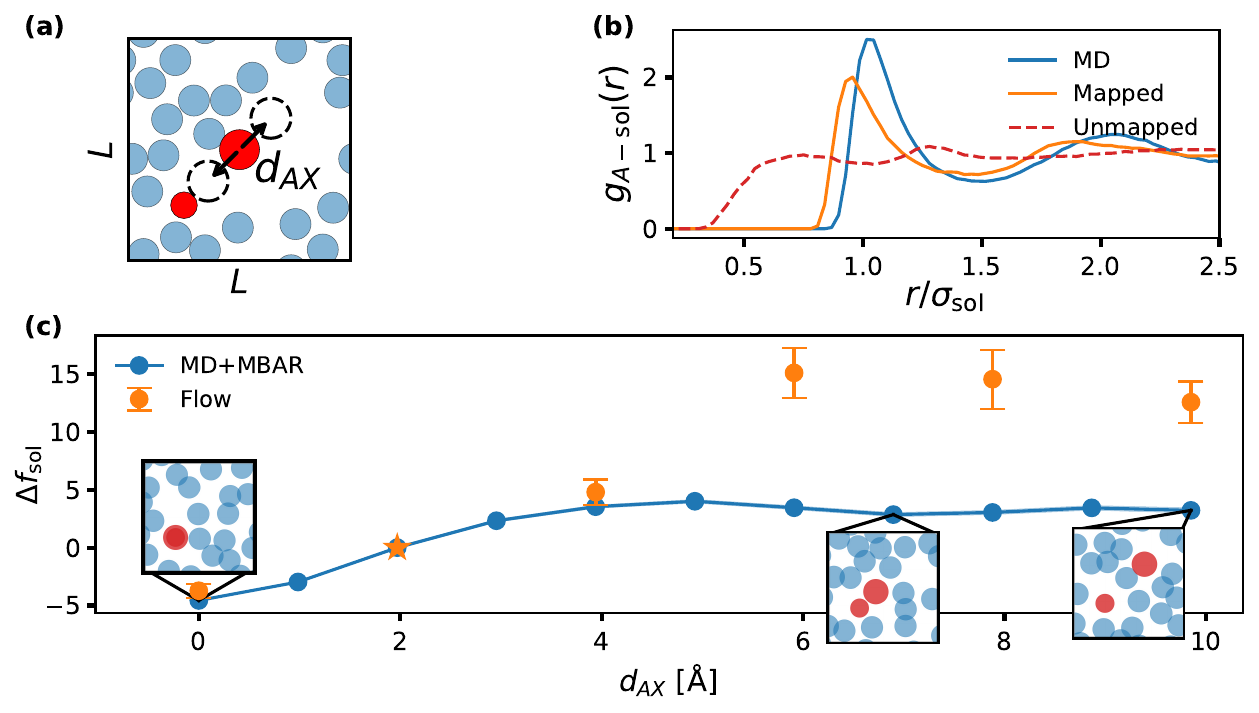}%
\caption{\label{fig:distance} (a) Illustration of the experiment in which the solute-solute-distance is varied. Solute and solvent particles are depicted as red and blue spheres, respectively. (b) RDF $g_{A-\mathrm{sol}}(r)$ between solute $B$ and solvent particles, showing both the unmapped and mapped RDFs compared to the reference MD RDF at a distance $d_{AX} = 4~$\AA. For the unmapped RDF, solvents configurations from the base state $d^0_{AX}=2.0~$\AA~ were combined with  the solutes placed at the target distance $d_{AX}=4.0~$\AA. 
(c) $\Delta f$ as a function of particle separation $d_{AX}$ relative to the base state $d^0_{AX}=2$~\AA, marked by a star symbol, estimated using both the flow method and MD+MBAR.}
\end{figure} 
As a second experiment, we vary the distance $d_{AX}$ between the two solute particles while setting $\sigma_A = 3.8$~\AA~ at $T=100$~K over a range of $d_{AX} = 0.0-10.0$~\AA. The base state is sampled at $d^0_{AX} = 2.0$~\AA~ and the flow needs to learn a mapping that reflects the rearrangement of the solvation shell as the two solute particles merge and separate, respectively.  Since the solute particles do not interact with each other, the change in free energy primarily arises from a change in the size of the cavity needed to accommodate both solutes. Intuitively, the free energy is expected to be higher at intermediate distances, where the accessible phase space for solvent particles is reduced. At large separations, the free energy is expected to plateau, since the solvation shells of the two solutes become uncorrelated and independent of their distance.
Figure~\ref{fig:distance}(c) shows the free‐energy difference relative to the base state at $d^0_{AX} = 2$~\AA~as a function of the distance between the non-interacting solutes. Around the base state, between $d_{AX} = 0-4$~\AA, the free-energy estimates produced by the flow model are in excellent agreement with MBAR values. In particular, the free energy increase up to a solute distance of about $d_{AX} = 5$~\AA~ is reproduced correctly, reflecting the decrease in accessible volume for solvent particles. For larger distances, the two $\Delta F$ profiles deviate noticeably, with the free energy estimates of the flow model being consistently too large.
Nevertheless, both methods predict a constant free energy difference at larger separations, indicating that the solvation shells no longer interact. 

The RDF between solute $A$ and the solvent particles is shown in Figure~\ref{fig:distance}(b) for a solute-solute distance of $d_{AX} = 4$~\AA, as obtained from MD simulations and compared to mapped and unmapped configurations from the base distribution.
In the unmapped case, the RDF is computed by combining solvent configurations from the base state ($d^0_{AX} = 2$~\AA) with the solutes placed at the target distance $d_{AX} = 4$~\AA. In the mapped case, the solvent configurations are updated using the flow to improve the alignment with the target solute placement at $d_{AX} = 4$~\AA. The RDF without any flow mapping of the solvent particles (red dashed curve in Fig.~\ref{fig:distance}(b)) does not exhibit any features  indicating the presence of  solvation shells, but rather shows a homogeneous distribution. It is also evident that solute and solvent particles are found too close to each other with distances as low as $r/\sigma_\text{sol} = 0.5$. Applying the flow mapping (orange curve in Fig.~\ref{fig:distance}(b)) recovers the typical peak structure of the RDF signifying the first and second solvation shell. 
Even though some discrepancies in peak position and height are observed between the mapped and MD results, the flow mapping substantially improves the overlap in configurations space, thereby indicating accelerated convergence of the free energy estimates  (see also Sec.~\ref{sec:flow_vs_mbar}).

\subsection{Computational efficiency}\label{sec:flow_vs_mbar}
\begin{figure}
\includegraphics[width=1\textwidth]{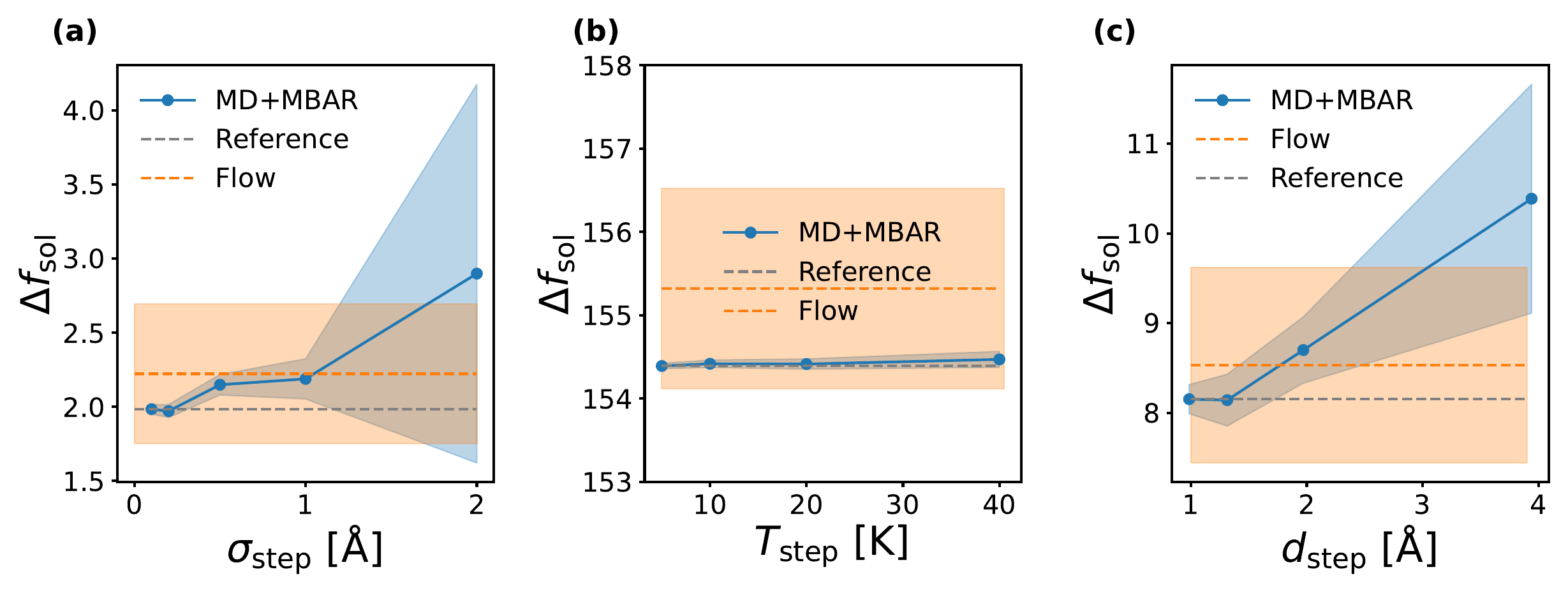}%
\caption{\label{fig:convergence} (a) $\Delta f_{\text{sol}}$ between two states with $\sigma_B = 2.0$~\AA~ and $\sigma_B$ = 4.0~\AA~ at $T = 100$~K as obtained from MD+MBAR shown as a function of the spacing between intermediate states $\sigma_{\rm step}$. (b)  $\Delta f_\mathrm{sol}$  as a function of the temperature step size between states with $\sigma_B =3.0$~\AA~ at $T=100$~K and 140~K. (c) $\Delta f_\mathrm{sol}$ as a function of the displacement step size for the transition from $d_{AX} = 0$~\AA~ to $d_{AX} = 4$~\AA~ at $T=100$~K.}
\end{figure}
The computational cost of free energy estimates is primarily dominated by the number of required energy and force evaluations. In the traditional MD+MBAR approach, this depends on the number of samples $S_{\rm MBAR }$ generated in each ensemble  and the number of intermediate ensembles between the two states. The $S_{\rm MBAR }$ samples need to be uncorrelated, meaning that only every $n$th configuration from an MD or MC trajectory can be used, with typical values of $n \approx 1000$.
For a flow-based approach with an MD generated base distribution, sampling $S^{\rm train}_{\rm BG}+S^{\rm eval}_{\rm BG}$ uncorrelated configurations for the base constitutes the major computational effort in terms of energy evaluations. The training of the flow model as well as the evaluation of Eq.~\eqref{eq:tfep} during inference require additional energy evaluations but significantly less than the sampling of the base distribution.
The number of samples needed in each MBAR ensemble and for the flow base distribution  depends on several factors, for example, in MBAR the overlap between and, therefore, number of intermediate states. For the following discussion of computational efficiency, we assume that a trained model is available, concentrating on the number of ensembles that must be sampled with MD in each case.

To compare the convergence of the free energy estimates within the MD+MBAR and flow-based approaches, we compute the MBAR estimator in Eq.~\eqref{eq:mbar} with different numbers of intermediate states for the three cases discussed for the solute/solvent system: changing the size of solute $B$, changing the temperature, and changing the distance $d_{AX}$ between solutes.
In Figure~\ref{fig:convergence}(a), the convergence of the free energy difference for changing the solute radius from $\sigma_B = 2$~\AA~ to $4$~\AA~ at $T=100$~K is shown  as a function of the spacing between intermediate states in the MBAR estimate. Only for a step size of around $1$~\AA, the MBAR estimates converge to an accuracy comparable to that of the flow-based predictions. This implies that, in addition to simulations at the end points, at least one intermediate state is required, resulting in a total of three simulated ensembles to ensure sufficient overlap between neighboring distributions.
Here, the flow model is more efficient as only the base distribution needs to be sampled and a direct mapping to different solute sizes is learned.

Computing the change in free energy as a function of temperature appears to be relatively robust within the MBAR approach. The free energy difference between $T=100$~K and 140~K for $\sigma_B = 3$~\AA~ is shown as a function of step width between intermediate states in Figure~\ref{fig:convergence}(b). Even for $\Delta T =40$~K, that is taking samples from the two end-states at $T=100$~K and 140~K only, the MBAR estimate achieves sufficient accuracy with a relatively small variance, even better than the estimate from the flow model. This exemplifies the strength of the MBAR estimator mixing contributions from both distributions as compared to TFEP used with the flow model, where the mapped samples are only evaluated in the target distribution. In this case, the flow model does not provide any pronounced gain in computational efficiency.  

Capturing the rearrangement of the solvent particles with increasing solute separation is challenging for both MD+MBAR and the flow model. 
The MBAR estimate for the free energy difference between $d_{AX} = 0$~\AA~ and 4~\AA~ is shown in Figure~\ref{fig:convergence}(c) as a function of spacing between intermediate states. For small step sizes of $\Delta d_{AX} \leq 1.5$~\AA, resulting in $2-3$ intermediate states in addition to the two end states, highly accurate values are obtained. Even with only a single intermediate state, that is  $\Delta d_{AX} =2$~\AA, the accuracy is still comparable to the flow model, whereas including only the end states into the MBAR estimator is not sufficient. 
Similar to the results shown in Figure~\ref{fig:convergence}(a), this indicates that three simulations are required for the MBAR calculations to achieve the accuracy of the flow-based predictions obtained from a single simulated ensemble.

\section{Discussion}
In this work, we demonstrate that free energy estimation using normalizing flows has the potential to become a viable and effective tool for accelerating solvation free energy calculations. In particular, we illustrate that the flow models can deliver accurate free energy differences for challenging transformations, such as increasing the solute–solute distance or changing the solute radius -- scenarios that typically require the sampling of multiple intermediate ensembles along the transformation to achieve convergence with MBAR. By leveraging the mapping capabilities of the flow to bridge between distributions, the required number of energy evaluations can be substantially reduced.

There are, however, limitations in the current implementation. Our study employed relatively small systems with simple model potentials, and further validation on more complex and realistic molecular systems remains to be explored. The range that can be reliably bridged by the flow is finite, as the complexity of the learned transformations increases, challenging the expressivity of the flow architecture. For small state spacings, the cost of generating the training samples may offset the efficiency gains provided by the mapping procedure.
A key bottleneck is that the current architecture updates solvent particles solely based on their absolute coordinates, without capturing correlations among them. We experimented with more expressive architectures, such as transformers,~\cite{vaswani2017attention} but did not observe improvements. This limitation reflects a broader challenge: existing flow models still struggle to fully capture the collective, many-body structure of liquids.~\cite{coretti_learning, schebek2024efficient} Potential architectural improvements that better incorporate solvent–solvent correlations  are expected to profoundly enhance the effectiveness of the method presented here, and will likely be even more important when extending the approach to more complex molecular systems.

\begin{acknowledgments}
This research has been funded 
by Deutsche Forschungsgemeinschaft (DFG) through grant CRC~1114 ``Scaling Cascades in Complex Systems'' (project number 235221301, Project B08  ``Multiscale Boltzmann Generators'' and B05 ``Origin of scaling cascades in protein dynamics'') and through CRC 1449 ``Dynamic Hydrogels at Biointerfaces'' (project number 431232613, Project C02 ``Rationally designed mucin-like glyco and peptide hydrogels''). The Flatiron Institute is a division of the Simons foundation.
\end{acknowledgments}

\newpage

\bibliography{article_clean}

\setcounter{table}{0}
\setcounter{figure}{0}
\renewcommand{\thefigure}{S\arabic{figure}}
\renewcommand{\theequation}{S\arabic{equation}}
\renewcommand{\thetable}{S\arabic{table}}
\setcounter{equation}{0}
\setcounter{table}{0}
\setcounter{section}{0}
\renewcommand\thesection{\Alph{section}}

\newpage
\section*{Supplementary Information}

\section{Model architectures and Training}
The flow architecture used in this work is based on the design introduced in Ref.~\onlinecite{wirnsberger_lfep}. Each layer $L$ of the coupling flow splits the system into two partitions of coordinates, $\bx = (\bx_A, \bx_B)$, and updates one partition conditioned on the other:
\begin{align}
\bx_A' &= f^L_{\theta}(\bx_A \,|\, g(\bx_B)), \\
\bx_B' &= \bx_B,
\end{align}
where $f^L_\theta$ is the transformation applied to $\bx_A$ and $g(\bx_B)$ computes features from the unchanged coordinates $\bx_B$ that condition the transformation. This splitting is performed along Cartesian coordinates, and each layer updates the corresponding coordinates of all particles. To ensure that every coordinate can be updated as a function of all other coordinates, we cycle through the partitions $\{1\}, \{2\}, \{3\}, \{1,2\}, \{2,3\}, \{1,3\}$ across layers. The invariant coordinates are projected onto a torus to ensure periodicity before feeding them into the conditioner $g$, which simplifies the learning task. For $g$, we tested different conditioners, including an attention-based variant~\cite{vaswani2017attention} and one using only absolute coordinates projected onto the torus. Since both yielded comparable results, we adopt the simpler, non-interacting conditioner. The transformation $f^L_{\theta}$ is implemented using circular splines.~\cite{rezende2020normalizingflowstorispheres, durkan2019neural} The flow models were optimized using the Adam optimizer~\cite{kingma_adam} with a learning rate of $7\cdot10^{-5}$ and trained on $S^{\rm train}_{\rm BG}=50{,}000$ samples drawn from the base distribution. For all experiments, convergence was reached quickly within a few tens of epochs.

\section{MD+MBAR simulation details}
All MD simulations were performed with OpenMM on the Curta HPC cluster at Freie Universität Berlin.~\cite{OpenMM,Bennett2020} 
Each system contained 100 particles, including two  solute particles with distinct interaction parameters. Simulations were run in a cubic periodic box ($L = 16.9~$\AA). The resulting reduced particle density of the system is therefore $\rho^*=0.8$. The particles interacted via a Lennard-Jones potential
\begin{align}
    V_\mathrm{LJ}(r_{ij}) = 4 \epsilon_{ij}\left ( \left (\frac{\sigma_{ij}}{r_{ij}} \right )^{12} - \left (\frac{\sigma_{ij}}{r_{ij}} \right )^{6} \right )\quad,
\end{align}
where $r_{ij}$ denotes the distance between two particles $i$ and $j$.
The interaction parameters $\sigma_{ij}$ and $\epsilon_{ij}$ are being determined from the parameters of the individual particles $\sigma_i$ and $\varepsilon_i$ using the Lorentz-Berthelot combining rule
\begin{align}
    \sigma_{ij} = \frac{\sigma_i+\sigma_j}{2}\quad,\\
    \epsilon_{ij} = \sqrt{\epsilon_i\epsilon_j}\quad.
\end{align}
We used a potential cutoff of $r_{\mathrm{cutoff}}=8$~\AA~and a switching distance of $r_{\mathrm{switch}}=5.0$~\AA~to ensure a continuous decay of the potential. The switching function $s(x)$ is multiplied with the energy, and is defined as~\cite{OpenMM}
\begin{align}
    s(x) = 1-6x^5+15x^4-10x^3\quad,
\end{align}
where $x$ is given by
\begin{align}
    x(r_{ij}) = \frac{r_{ij}-r_{\mathrm{switch}}}{r_{\mathrm{cutoff}}-r_{\mathrm{switch}}}\quad.
\end{align}
The interaction lengths of the solutes were set to $\sigma_X = 3.0$~\AA~ and $\sigma_A = 3.8$~\AA, unless specified differently, and  $\sigma_{\mathrm{sol}}=3.4$~\AA~ for all solvent particles. All particles shared the same energy scale $\epsilon_i = 0.996~$kJ/mol. The interaction potential between the two solute particles was set to $V(r_{XY}) = 0$. Solvent masses were set to $m_{\mathrm{sol}}=39.9~$amu and the solutes were position-constrained. Initial coordinates were placed on a grid with spacing $d_{\mathrm{space}} = 3.6~$\AA. The system was minimized using L-BFGS followed by 5,000 steps of gradient descent. Thermalization was carried out with a Nos{\'e}–Hoover thermostat for 50,000 steps at the target temperature and a timestep of $\Delta t =0.01~$fs, ensuring that all samples had a similar and initially stable starting configuration. Since the production run has to be simulated with a stochastic integrator to ensure uncorrelated samples, a Langevin thermostat was used to further  thermalize the system for 50,000~fs at $\Delta t =0.1~$fs. Production simulations were run for 10.5~ns using $\Delta t=1.5~$fs with a Langevin thermostat with a coupling time of $\tau = 0.2~$ps. Configurations were saved every 7,000 steps.\\

\end{document}